\def\ga{\mathrel{\mathpalette\fun >}}
\def\fun#1#2{\lower3.6pt\vbox{\baselineskip0pt\lineskip.9pt
  \ialign{$\mathsurround=0pt#1\hfil##\hfil$\crcr#2\crcr\sim\crcr}}}
\relax
\documentstyle[12pt]{article}
\advance\textheight by \topskip
\begin{document}
\thispagestyle{empty}

\phantom{.}
\rightline{SU-ITP-94-37}
\rightline{astro-ph/9410039}
\rightline{October 12, 1994}
\vskip 0.5cm

\begin{center}
\Large
{\bf Stationary Universe Model:  Inputs and Outputs}\footnote{Invited
talk
given at the Rome Conference ``Birth of the Universe and Fundamental
Physics'',
May,1994.} \\
\vskip 1.0cm

{\sc Arthur Mezhlumian}\footnote{On leave from Landau Institute for
Theoretical
Physics, Moscow, Russia.}\\
\vskip 0.5cm

{\large \em
Department of Physics, Stanford University \\
Stanford, CA 94305-4060, USA}\footnote{E-mail:
arthur@physics.stanford.edu}\\
\end{center}
\vskip 0.5cm

\def\lsim{\mathrel{\lower2.5pt\vbox{\lineskip=0pt\baselineskip=0pt
           \hbox{$<$}\hbox{$\sim$}}}}
\def\gsim{\mathrel{\lower2.5pt\vbox{\lineskip=0pt\baselineskip=0pt
           \hbox{$>$}\hbox{$\sim$}}}}
\def\bel#1{\begin{equation}\label{#1}}
\def\eel#1{\label{#1}\end{equation}}

\newcommand{\DD}[2]{\mbox{$\frac{\partial^{#1}}{\partial #2}$}}
\newcommand{\Ddd}[2]{\mbox{$\frac{\partial^{2}}{\partial #1 \partial
#2}$}}
\newcommand{\uv}[2]{\mbox{$u(#1,#2,z,{\cal D})$}}
\newcommand{\Dme}{\mbox{$m_{1}(t,{\cal D}|\chi)$}}
\newcommand{\Dma}[1]{\mbox{$m_{1}(t,{\cal D_{#1}}|\chi_{#1})$}}
\newcommand{\EV}[1]{\mbox{${\bf{\sf E}}\left\{ \left. #1 \right| \,
\phi_0 =
\chi \right\} $}}   \newcommand{\Pes}[1]{\mbox{$\psi_{#1}(\chi) $}}
\newcommand{\Dpi}[1]{\mbox{$\pi_{#1}(\phi) $}}
\newcommand{\tW}[1]{\mbox{$\tilde{W}^{#1}(\phi) $}}
\newcommand{\TW}[1]{\mbox{$\tilde{W}^{#1}(\phi(z)) $}}
\newcommand{\Tw}[1]{\mbox{$\tilde{W}^{#1}(z) $}}

\noindent  This talk presents the recent progress achieved in
collaboration
with
A.Linde and D.Linde$^{\ref{MezhLin},\ref{LLMPascos},\ref{Omega}}$
towards
understanding the true nature of the
global spatial structure of the universe as well as the most general
stationary
characteristics of its time-dependent state with eternally growing
total
volume.

In our opinion, the simplest and, simultaneously, the most
general  version of  inflationary cosmology  is the chaotic inflation
scenario.   It can be realized in all models were the other
versions of inflationary theory can be realized
(see [\ref{LindeBook}] for detailed account).
Several years ago it was
realized that inflation in these theories has a very interesting
property$^{\ref{b19},\ref{b20}}$ which  will be discussed in this
talk.
If the universe contains at least one inflationary domain of a size
of horizon
($h$-region) with a sufficiently large and homogeneous scalar field
$\phi$, then
this domain will permanently produce new $h$-regions of a similar
type. During
this process the total physical volume of the inflationary universe
(which is
proportional to the total number of $h$-regions) will grow
indefinitely.

 Fortunately, some
kind of stationarity may exist in many models of  inflationary
universe due to
the process of the universe self-reproduction$^{\ref{LindeBook}}$.
The
properties of
inflationary domains formed during the process of the
self-reproduction of the
universe do not depend on the moment of time at which each such
domain is
formed; they depend only on the value of the scalar fields inside
each domain, on
the average density of matter  in this domain and on the physical
length scale. This talk will describe what assumptions and
approximations
are made in order to get this picture of the universe and what
consequences they bear.

Let us consider the simplest
model of chaotic inflation based
on the  theory of a  scalar field $\phi$ minimally coupled to
gravity, with the
Lagrangian
\begin{equation}\label{E01}
L =  \frac{1}{16\pi}R + \frac{1}{2} \partial_{\mu} \phi
\partial^{\mu} \phi - V(\phi)  \ .
\end{equation}
Here $G = M^{-2}_p = 1$ is the gravitational constant,  $R$ is the
curvature
scalar, $V(\phi)$ is the effective potential of the scalar field.
The most important fact for inflationary scenario is that for most
potentials $V(\phi)$ (e.g., in all power-law $V(\phi)=g_n \phi^n /n$
and
exponential $V(\phi)=g e^{\alpha \phi}$ potentials) there is an
intermediate
asymptotic regime of slow rolling of the field $\phi$ and
quasi-exponential
expansion of the universe. This expansion (inflation)
ends at $\phi \sim \phi_e$ where the slow-rolling regime $\ddot\phi
\ll
3H(\phi) \dot\phi$ breaks down.

During the inflation all the inhomogeneities are stretched away and,
if the
evolution of the universe were governed solely by classical equations
of
motion,  we would end up with extremely smooth
geometry of the spatial section of the universe with no primordial
fluctuations to initiate the growth of galaxies and large-scale
structure.
Fortunately, the same gravitational instability which causes the
growth of
galaxies
during the Hot Big Bang era leads to the existence of the growing
modes of
vacuum fluctuations during the inflation. The wavelengths of all
vacuum
fluctuations of the scalar field $\phi$ grow exponentially
in the expanding universe. When the wavelength of any
particular fluctuation becomes greater than $H^{-1}$, this
fluctuation stops oscillating, and its amplitude freezes at
some nonzero value $\delta\phi (x)$ because of the large
friction term $3H\dot{\phi}$ in the equation of motion of the field
$\phi$\@. The amplitude of this fluctuation then remains
almost unchanged for a very long time, whereas its
wavelength grows exponentially. Therefore, the appearance of
such a frozen fluctuation is equivalent to the appearance of
a classical field $\delta\phi (x)$ that does not vanish
after averaging over macroscopic intervals of space and
time.

Because the vacuum contains fluctuations of all
wavelengths, inflation leads to the creation of more and
more perturbations of the classical field with
wavelengths greater than $H^{-1}$\@. The average amplitude of
such non-linearly modulated perturbations
generated during an e-fold time interval $H^{-1}$ is given by
\begin{equation}\label{E23}
|\delta\phi(x)| \approx \frac{H}{2\pi}\ .
\end{equation}
 If the field is effectively massless (i.e.\  $m^2(\phi) \ll
H^2(\phi)$), the amplitude of each frozen wave does not change in
time.
On the other hand, phases of each waves are random.
Therefore,  the sum of all waves at a given point fluctuates and
experiences
Brownian jumps in all directions in the space of fields.

The standard way of description of the stochastic behavior of the
inflaton
field during the slow-rolling stage is to coarse-grain it over
$h$-regions and
consider the effective equation of motion of the long-wavelength
field$^{\ref{b60},\ref{b61},\ref{Star}}$:
 \begin{equation} \label{m1}
\frac{d}{dt} \, \phi = - \frac{V'(\phi)}{3H(\phi)} +
\frac{H^{3/2}(\phi)}{2\pi}
\, \xi(t) \ ,
\end{equation}
Here $\xi(t)$ is the effective white noise generated by quantum
fluctuations,
which leads to the Brownian motion of the classical field $\phi$\@.

This Langevin equation corresponds to the following Fokker-Planck
equation
for the probability density $P_c(\phi,t)$ to find the field $\phi$ at
a
given point (which now means $h$-region) after time $t$:
\begin{equation}\label{E3711}
\frac{\partial P_c}{\partial t} =
\frac{\partial}{\partial \phi} \left({H^{3/2}(\phi)\over
8\pi^2} \
\frac{\partial}{\partial \phi}\, \left(H^{3/2}(\phi) P_c \right) +
{V'(\phi)\over 3H(\phi)} \, P_c\right) \ ,
\end{equation}

The formal stationary solution ($\partial P_c/\partial t=0$) of
equation (\ref{E3711}) would be$^{\ref{b60}}$
\begin{equation}\label{E38}
P_c \sim \exp\left({3\over 8 V(\phi)}\right) \ ,
\end{equation}

Note, that the stationary solution (\ref{E38}) is equal to the square
of the
Hartle-Hawking wave function of the universe$^{\ref{HH}}$.  At  first
glance, this
result is a direct confirmation of the Hartle-Hawking
prescription for the wave function of the universe.

However, in all realistic cosmological theories, in which $V(\phi)=0$
at its
minimum, the distribution (\ref{E38}) is not
normalizable. The source of this difficulty can be easily
understood: any stationary distribution may exist only due
to a compensation of a classical flow of the field $\phi$
downwards to the minimum of $V(\phi)$ by the diffusion motion
upwards. However, diffusion of the field $\phi$ discussed
above exists only during inflation, i.e. only for $\phi \geq
1$, $V(\phi)\geq V(1)\sim m^2\sim 10^{-12}$ for $m \sim 10^{-6}$.
Therefore (\ref{E38}) would correctly describe the stationary
distribution $P_c(\phi)$ in the inflationary universe only if
$V(\phi)\geq 10^{-12} \sim 10^{80}$ GeV in the absolute minimum of
$V(\phi)$, which is, of course, absolutely unrealistic$^{\ref{b20}}$.

It can be shown$^{\ref{MezhLin}, \ref{LLMPascos}, \ref{b20}}$
that the solutions of this equation with the effect
of ``end of inflation'' boundary properly taken into account are
non-stationary (decaying). It is of no surprise because we didn't
take into
account yet the complicated and inhomogeneous expansion of the whole
universe with multiplicating number of $h$-regions.

In order to describe the structure of the inflationary universe
beyond one
$h$-region (minisuperspace) approach one has to investigate the
probability
distribution $P_p(\phi,t)$, which  takes into account the
inhomogeneous
exponential growth of the volume of domains filled by field
$\phi$$^{\ref{b19},\ref{b20}}$.  The close view on the
process of the growth of the volume of inflationary universe reveals
the
strong resemblance with branching diffusion processes$^{
\ref{MezhLin},\ref{LLMPascos},\ref{MezhMolch}}$, with the role of
branching particles being played by $h$-regions and the the diffusing
parameter being associated with the inflaton field inside an
$h$-region.

A branching diffusion process is characterized by the independent
diffusion of particles within the allowed region and their splitting
at
random times into several ``daughter'' particles which then continue
to
diffuse independedly from their birthplace. The characteristic
splitting
time is related to the intensity of branching which, in addition to
diffusion
coefficient and drift term of ordinary diffusion, determines the time
evolution
of the distribution of particles in the space where they live.

To describe the distribution of the inflaton field in the whole
universe and not only in one Hubble domain, one has to modify the
Fokker-Planck equation by introducing the branching intensity term
corresponding to creation of
more and more new ``particles'' during the diffusion process$^{
\ref{MezhLin},\ref{LLMPascos},\ref{ArVil},\ref{Nambu},\ref{Mijic},\ref
{ZelLin}}$\@.
\begin{equation}\label{E372}
\frac{\partial P_p}{\partial t} = \frac{\partial }{\partial\phi}
 \left( {H^{3/2}(\phi)\over 8\pi^2} \frac{\partial }{\partial\phi}
\left(
 {H^{3/2}(\phi)}P_p \right)
 +  \frac{V'(\phi)}{3H(\phi)} \, P_p \right)
 +  3H(\phi)  P_p
\end{equation}

The mathematical model describing such behavior is
the recently developed theory of a new type of branching diffusion
processes$^{\ref{MezhMolch}}$ where ``new'' refers to diffusion
taking
place in the parameter space rather than the space where the
particles
live. Indeed, in our case the $h$-regions live in ordinary space,
while
the parameter associated with each $h$-region is the value of the
coarse-grained inflaton field in it and this parameter lives on the
space
of homogeneous fields which is in the simplest case a segment of the
real line $\left[ \phi_e, \phi_p \right]$\@.  Eq.\ (\ref{E372}) is,
from this
perspective, the forward Kolmogorov equation for the first moment of
the
generating functional of number of branching particles.

There are two main sets of questions which may be asked concerning
such
processes. First of all, one may
be interested in the probability $P_p(\phi, t)$ to find a given field
$\phi$ at a
given time $t$ under the condition that initial value of the field
was equal to
some $\phi(t=0) = \phi_0$\@. In what follows we will denote $\phi_0$
as $\chi$\@. On the other hand, one may wish to know, what is the
probability
$P_p(\chi,t)$  that the given final value of the field $\phi$ (or the
state
with a given final density $\rho$) appeared as a process of diffusion
and
branching of a domain containing some field $\chi$\@. Or, more
generally, what
is the typical history of a branching Brownian trajectory which ends
up at a
hypersurface of a given $\phi$ (or a given $\rho$)?

It happens that if we do not take the Planck boundary seriously, this
equation also doesn't have a true stationary solution (the solutions
found
by Nambu {\it et.\ al.}$^{\ref{Nambu}}$ are heavily concentrated at
the
super-Planckian densities which is just another way to state that
there is no
real stationarity). This conclusion, however, may fail if we will
treat the
Planck boundary more carefully. There are different reasons to do
this:
\begin{enumerate}
\item Diffusion equations were derived in the semiclassical
approximation for quantum scalar field which breaks down near the
Planck energy density $\rho_p \sim M^4_p =1$.
\item Interpretation of the processes described by these
equations is based on the notion of classical fields in a classical
space-time,
which is not applicable at densities larger than $\rho_p \sim 1$
because of large fluctuations of metric at such densities. In
particular, our
interpretation of $P_c$ and $P_p$ as of probabilities to find
classical field $\phi$ in a given point at a given time does not make
much
sense at $\rho > 1$.
\end{enumerate}

There is also another, rather intrinsic for inflationary model,
reason to expect the existence of the
stationary solutions: as we will argue now, inflation kills itself as
the
density approaches the Planck density $\rho_p \sim M_p^4$.

In our previous investigation we assumed that the vacuum energy
density is given by $V(\phi)$ and the energy-momentum tensor is given
by
$V(\phi)\, g_{\mu\nu}$\@. However, quantum fluctuations of the scalar
field give the contribution to the average value of the energy
momentum
tensor, which does not depend on mass (for $m^2 \ll H^2$) and is
given
by$^{\ref{b99}}$
\begin{equation}\label{x1}
 <T_{\mu\nu}> \, = \, {3\, H^4\over 32\, \pi^2}\ g_{\mu\nu} \, =\,
 {2 \over 3}\, V^2 \ g_{\mu\nu} \ .
\end{equation}
The origin of this contribution is obvious. Quantum
fluctuations of the scalar field $\phi$ freeze out with the amplitude
${H\over 2\pi}$ and the wavelength $\sim H^{-1}$\@. Thus, they lead
to
the gradient energy density $(\partial_\mu\delta\phi)^2\sim H^4$.

An interesting property of eq. (\ref{x1}) is that the average value
of the
energy-momentum tensor of quantum fluctuations does not look  like an
energy-momen\-tum tensor corresponding to the gradients  of a
sinusoidal wave.
It looks rather like a renormalization of the vacuum energy-momentum
tensor (it
is proportional to $g_{\mu\nu}$)\@. This means, in particular, that
after averaging over all possible outcomes of the process of
generation of
long-wave
perturbations, the result (for $<T_{\mu\nu}>$) does not depend on the
choice of
the coordinate system. However, if we are interested in local events
(i.e.\
we are averaging only over short wavelengths),
 we will see long-wave inhomogeneities produced by the ``frozen''
fluctuations of  the scalar field. This does not lead to any
interesting
effects at $V \ll 1$ ($\phi \ll \phi_p$)  since in this case $V^2 \ll
V$.

However, at the density larger than the Planck density  the situation
becomes much more complicated. At $V > 1$ the gradient energy density
$\sim
V^2$ becomes  larger than the potential energy density $V(\phi)$\@. A
typical
wavelength of perturbations giving the main contribution to the
gradient
energy is given by the size of the horizon, $l \sim H^{-1}$\@. This
means that the
inflationary universe at the Planck density becomes divided into many
domains
of the size of the horizon, density contrast between each of these
domains
being of the order of one. These domains evolve as separate
mini-universes with
the energy density dominated not by the potential energy density but
by the
 energy density of gradients of the field $\phi$\@. Such domains
drop out from the process of exponential expansion. Some of them may
re-enter
this process later, but many of them will collapse within the typical
time $H^{-1}$.

This can be effectively described by imposing some kind of absorbing
(or
reflecting, or elastic screen type) boundary conditions on our
process of
branching diffusion at some $\phi_b$ which should be close to the
Planck
boundary. Our calculations have shown$^{\ref{MezhLin}}$ that if the
boundary conditions do not permit the field $\phi$
penetrate deeply into the domain $\phi > \phi_p$,
the final results of our investigation do not depend on the
type of the boundary conditions imposed (whether they are absorbing,
reflecting, etc.)\@. They depend only on the value of the field
$\phi_b$
where the boundary condition is to be imposed, and this dependence is
rather trivial. We have argued that $\phi_b \sim \phi_p$\@.
Therefore we will assume now that the function $P_p(\phi,t|\chi)$
satisfies boundary conditions corresponding to disappearing particles
at
$\phi_p$ treating $\phi_p$ now just as a phenomenological parameter
not
necessarily corresponding to $V(\phi_p)=1$\@.

The boundary conditions at the ``end of inflation'' boundary can be
derived
from the requirement of the continuity of the probability and its
flux through
that boundary$^{\ref{MezhLin},\ref{LLMPascos}}$\@. Their form
suggests that
the dependence of the physical probability distribution on the value
of the
initial field $\chi$ near that boundary is close to square of the
``tunneling
wavefunction''$^{\ref{tunnel}}$.

We have shown$^{\ref{MezhLin},\ref{LLMPascos}}$ that the asymptotic
solution for $P_p(\phi,t|\chi)$ (in the limit $t \rightarrow \infty$)
is given
by
\begin{equation} \label{eq22}
P_p(\phi,t|\chi) \rightarrow e^{\lambda_1 t}\, \Pes{1} \,
\Dpi{1} \ .
\end{equation}
Here $\Pes{1}$ is the only positive eigenfunction of the equation:
\begin{equation} \label{eq15}
\frac{1}{2} \frac{H^{3/2}(\chi)}{2\pi} \frac{\partial}{\partial\chi}
 \left( \frac{H^{3/2}(\chi)}{2\pi} \frac{\partial }{\partial\chi}
\Pes{s} \right) - \frac{V'(\chi)}{3H(\chi)} \frac{\partial
}{\partial\chi}
\Pes{s}
+ 3H(\chi) \, \Pes{s} = \lambda_s \, \Pes{s}  \ .
\end{equation}
$\lambda_1$ is the corresponding (real) eigenvalue, and $\pi_1(\phi)$
(invariant density of branching diffusion) is the eigenfunction
of the conjugate operator with the
same eigenvalue $\lambda_1$\@. We found that
in realistic theories of inflation the typical time of
relaxation to the asymptotic regime is extremely
short. It is only about a few thousands Planck times, i.e. about
$10^{-40}~sec$\@. This means, that the normalized distribution
\begin{equation} \label{eq22aa}
\tilde{P}_p(\phi,t|\chi) = e^{-\lambda_1 t}
\,P_p(\phi,t|\chi)
\end{equation}
rapidly converges to a time-independent function:
\begin{equation} \label{eq22a}
\tilde{P}_p(\phi|\chi) \equiv
\tilde{P}_p(\phi,t \rightarrow \infty|\chi) =  \Pes{1} \, \Dpi{1} \ .
\end{equation}
It is this stationary distribution that we were looking for.
After some calculations we came to the following expression for
$P_p(\phi|\chi)$ (note that the functions $\Phi(\cdot)$ are
essentially the
same, the only difference is the argument):
\begin{equation} \label{eq32}
P_p(\phi|\chi) = \frac{C}{V^{3/4}(\phi)} \, \exp \left\{
\frac{3}{16\, V(\phi)}
- \frac{3}{16\, V(\chi)} \right\} \, \Phi(z(\chi)) \,
\Phi(z(\phi))
\end{equation}
Here $C$ is the normalization constant. Using the WKB approximation,
we have
calculated$^{\ref{MezhLin},\ref{LLMPascos}}$ $\Phi(z(\phi))$  for a
wide class
of potentials
usually considered in the context of chaotic inflation, including
potentials
$V \sim  \phi^n$ and $V \sim e^{\alpha\phi}$\@.

The solution we have found features interesting properties. First
of all, it is concentrated heavily at the highest allowed values
of the inflaton field. Despite the promising exponential prefactor
in (\ref{eq32}) which looks like a combination of the
Hartle-Hawking$^{\ref{HH}}$ and tunneling$^{\ref{tunnel}}$
wavefunctions, the dependence of $\Phi(z(\phi))$ on $\phi$ appeared
to be even
steeper that those exponents. This function falls exponentially (with
the
rate governed by the Planckian energies) towards the lower values of
the
inflaton field and strongly overwhelms the dependence of the exponent
in front
of it. Only near the ``end of inflation'' boundary the solution
(\ref{eq32})
reveals something familiar --- the dependence on the initial field
$\chi$
becomes similar to the square of tunneling wavefunction
(simultaneously, the
corresponding dependence on the final field $\phi$ cancels out)\@.
And, as we
have already mentioned, the relaxation time is very small (which is
also a
consequence of the fact that the dynamics of the self-reproducing
universe is
governed by the maximal possible energies)\@. The
stationary distribution found in$^{\ref{MezhLin},\ref{LLMPascos}}$ is
not very
sensitive to our  assumptions concerning the concrete mechanism of
suppression  of production  of inflationary domains with $\phi \ga
\phi_p$\@.   We hope that these results may show us a way towards the
complete
quantum mechanical description of
the stationary ground state of the universe.

At the first glance it seems that there can be no observable
consequences of
the stationarity of inflation and the picture of self-reproducing
universe
which we just described. Indeed, according to traditional
inflationary paradigm
as well as according to our understanding, all the visible universe
is produced
during the last stages of the inflaton field rollover, when its
energy is
already much below not only the Planck scale but also the scale
$V(\phi_*)$
where the quantum diffusion becomes more important than the classical
slow
rolling. Therefore it seems that the effects related to essentially
Planckian
energies should determine the structure of the universe at such
exponentially
large scales which will inevitably be far beyond the observational
limits, thus
leaving us with traditional description of the visible part of the
universe.

There is, however, one important fact which follows from stationarity
and which
should alert us about the possibility that we are overlooking a
non-trivial
effect. The very essence of stationary picture implies that the rate
of the
volume growth is constant and governed by the highest accessible
energies!
However, if we look only at the local (produced no more than 60
e-folds prior
to the end of inflation) part of the universe, we are forced to
conclude that
those e-folds were achieved at much slower rate corresponding to
nearly
end-of-inflation energies. The resolution of the controversy is in
the fact
that the high total rate of volume growth for most values of inflaton
field is
achieved not by direct local expansion but by expansion at the
highest
accessible energies and subsequent slow rollover.

In other words, one can say that once we started to talk about the
volume
weighted ``physical'' probability distribution $P_p(\phi, t|\chi)$,
we should
be consistent and consider other effects with same volume weighted
measure.
Then it should not be surprising that some events which are very
improbable in
usual measure become highly probable in this new measure.

For example, in the traditional inflationary paradigm one obtains the
typical
evolution of the inflaton field at the end of inflation to be given
by
classical slow rollover plus comparatively small random noise
(\ref{E23}). This
result is a consequence of considering only local evolution of the
field, where
any deviations from it are strongly suppressed in the measure
corresponding to
$P_c(\phi, t)$ which can be written$^{\ref{LindeBook}}$ as
$\exp{\left(-\frac{2 \pi^2 \, (\delta \phi)^2}{H^2(\phi)}\right)}$\@.
However,
if we look at the problem from the volume weighted point of view, we
discover
that the typical trajectory of the inflaton is the one which
maximizes the
overall volume of the universe. In other words, although some
trajectory may be
suppressed by the local probability, if it gives large enough
contribution to
the growth of the volume of the universe, one will find many such
trajectories
in the unit of physical volume$^{\ref{Omega}}$.

Let us find the typical inflaton trajectory in volume weighted
measure. If it
was typical in this measure for inflaton to sit at highest energies
up to the
time when it has to start rolling down in order to approach the end
of
inflation by the time of our observation, it might be even better for
it to sit
there for a little longer and then to roll down with slightly higher
speed than
it is suggested by the classical slow rollover, because this will
enable the
part of the universe under consideration to expand even longer with
higher rate
and to produce more volume. However, the only way to achieve rollover
speeds
greater than the classical is for the quantum fluctuations to add up
with one
sign to greater jumps than it is suggested by (\ref{E23})\@.

Let the extra time interval spent at highest energies be
$\tilde{\Delta t}$\@.
Then we win the volume by factor of $e^{3 H_{max} \tilde{\Delta t}
}$\@.
However, to compensate for the lost time the inflaton has to jump at
least once
with the amplitude $ \tilde{\delta \phi} $ such that:
\begin{equation}\label{deltat}
\tilde{\Delta t}(\phi) = \frac{\tilde{\delta \phi}}{\dot{\phi}} =
\frac{ n(\phi) \frac{H(\phi)}{2 \pi}}{\dot{\phi}} = n(\phi) \, \Delta
t(\phi)
\end{equation}
where we introduced the factor $n(\phi)$ by which the jump is
amplified. Such
jump is suppressed in probability by the factor $e^{-\frac{1}{2} \,
n^2(\phi)}$\@. The jump at a given value of $\phi$ occurs with such
amplitude
that maximizes the volume weighted probability:
\begin{equation}\label{maximize}
\max{\left[\exp{\left(3 H_{max} n(\phi) \Delta t - \frac{1}{2}
n^2(\phi)\right)}\right]}
\end{equation}
which gives the answer
\begin{equation}\label{maximal}
n(\phi) = 3 H_{max}  \Delta t(\phi)
\end{equation}
In fact, we have found$^{\ref{Omega}}$ that the typical trajectory
consists
entirely of such subsequent jumps. We can relate the amplitude of the
jump to
the characteristics of the universe at a given length scale through
the
field-dependent value of the time
$\Delta t(\phi) = \delta \phi / \, \dot{\phi}$ of the conventional
quantum
jumps.

Recall the expressions for the amplitudes of scalar and tensor
perturbations
generated at given inflaton field   $\phi$ and therefore associated
with a
length scale through the usual slow rolling:
\begin{equation}\label{perturbscalar}
A^{pert}_S(\phi) = \left( \frac{\delta \rho}{\rho} \right)_S =
\left. c_S \, \frac{H(\phi) \delta \phi}{\dot{\phi}} \right|_{k \sim
H}
\end{equation}
\begin{equation}\label{perturbtensor}
A^{pert}_T(\phi) = \left( \frac{\delta \rho}{\rho} \right)_T =
\left. c_T \, \frac{H(\phi) }{M_p} \right|_{k \sim H}
\end{equation}
from where one obtains:
\begin{equation}\label{scalarovertensor}
\frac{A^{pert}_S(\phi)}{A^{pert}_T(\phi)} = \frac{c_S}{c_T} \, M_p \,
\frac{\delta \phi}{\dot{\phi}}
= \frac{c_S}{c_T} \, M_p \, \Delta t (\phi)
\end{equation}

Using this result we can rewrite (\ref{maximal}) in the form:
\begin{equation}\label{maximal2}
n(\phi) = 3 \, \frac{H_{max} }{M_p} \, \frac{c_T}{c_S} \,
\frac{A^{pert}_S(\phi)}{A^{pert}_T(\phi)}
\end{equation}

In the same way as the conventional amplitude of jumps $H/2 \pi$ is
related
with the well known perturbations of the background energy density
the
``nonperturbatively amplified'' jumps which we have just described
are related
with ``nonperturbative'' contribution to deviations of the background
energy
density from its average value. The correct interpretation of this
result is
that at the length scale associated with the value of the field
$\phi$ there is
an additional  nonperturbative contribution to {\it monopole}
amplitude:
\begin{equation}\label{nonperturb}
A^{nonpert}_S(\phi) = \left(
\frac{3 c_T}{c_S} \,   \frac{H_{max} }{M_p} \,
\frac{A^{pert}_S(\phi)}{A^{pert}_T(\phi)}
\right) \, A^{pert}_S
\end{equation}

The reason why there is no contribution to higher multipoles is that
large
quantum jumps  proceed through formation of spherically symmetric
transition
regions (all deviations from spherical symmetry are suppressed
exponentially).
However, one could justifiably worry about the induced dipole
contribution
which might come from the observation of a spherically symmetric
distribution
from a non-central point. We have found$^{\ref{Omega}}$ reasonable
constraints
on such induced contribution.

The self-consistency constraints for these results are
found$^{\ref{Omega}}$ to
be:
\begin{equation}\label{constraints1}
n(\phi) \gg 1 \; \Longleftrightarrow \; \frac{A^{pert}_S}{A^{pert}_T}
\gg 1
\end{equation}
\begin{equation}\label{constraints2}
n(\phi) H(\phi) \ll M_p \; \Longleftrightarrow \; (\nabla \phi)^2 \ll
V(\phi)
\end{equation}
which do not lead to any new restrictions for all models which are
otherwise
good from the observational point of view. There is another
constraint which is
not mentioned here and which can sometimes become important,
especially in the
cases when $\frac{A^{pert}_S}{A^{pert}_T} $ gets extremely large.

Since at any scale the nonperturbative contribution is spherically
symmetric
(monopole-like) and the sign of this contribution is definite (the
jumps of
inflaton field are definitely towards the lower values) one can argue
that
observationally such a configuration will look like a void with
scale-dependent
density which is lower than the density of the surroundings. Such a
void will
resembe in many aspects a locally open universe. Indeed, if there is
a
non-perturbative spherically symmetric hill in the Newtonian
potential and we
are near the top of that hill, we should see additional spherically
symmetric
radial component of peculiar velocities at which the galaxies are
moving from
us at farther distances as compared to the uniform Hubble law of flat
universe
--- a picture not very different from the open universe, at least in
what
concerns the motion of the matter. However, it will be interesting to
obtain
other observational characteristics of such spatial distribution of
Newtonian
potential.

The amplification factor in (\ref{nonperturb}) is small unless
$H_{max} \sim
M_p$, which tells us that one indeed needs to rely on the possibility
of
inflation going as high as Planckian energies in order to get such an
effect.
On the other hand it is proportional to the ratio of the scalar
perturbative
amplitude to tensor perturbative amplitude, which is  a quite large
number in
all reasonable inflationary theories, and may become very large in
some
versions of inflation with especially low end-of-inflation energy
scale. For
example, in the usual $\lambda \phi^4$ chaotic inflation one gets
$\frac{A^{pert}_S}{A^{pert}_T} \sim 10$, $H_{max} \sim M_p$, and
finally
$A^{nonpert}_S \sim 40 \, A^{pert}_S \sim 10^{-3}$\@.

In some more exotic inflationary scenarios one can get a huge
amplification
factor  $\frac{A^{pert}_S}{A^{pert}_T} $ and our non-perturbative
monopole
contribution may actually become of order $ 1$\@. This would be the
best so far
way to reconcile $\Omega < 1$ with inflation retaining small
anisothropy and
other well established achievements of inflation. However, this is a
case when
one should be very careful about checking the compliance with
self-consistency
constraints which may invalidate the result. There are at least some
scenarios
which survive the first crude check and we are currently working  on
refining
them.

As we promised in the title of this talk, we have outlined the inputs
which go
into building the stationary universe scenario, and quite
surprisingly we have
found some observational outputs which are so unique and unmistakable
that they
would be very difficult to mimic in any other model. They
intrinsically depend
on the assumption that the inflation can go up to Planckian energies
and that
it continues essentially forever. It becomes apparent that the
picture of
stationary universe is not only theoretically consistent and
aesthetically
appealing but is also testable and might be just the one needed to
describe the
observational data.

\vskip 0.5cm
\centering{\Large \bf References}
\begin{enumerate}
\item \label{MezhLin} A.D. Linde and A. Mezhlumian,  {\it Phys.\
Lett.}
B {\bf 307},  25  (1993).
\item \label{LLMPascos} A.D. Linde, D.A. Linde, and
A. Mezhlumian,  {\it Phys.\   Rev.} D {\bf 49},  1783  (1994).
\item \label{Omega} A.D. Linde, D.A. Linde, and A. Mezhlumian, {\it
to appear}.
\item \label{LindeBook} A.D. Linde, {\bf Particle Physics and
Inflationary
Cosmology} (Harwood, Chur, Swit\-zerland, 1990);
A.D. Linde, {\bf Inflation and Quantum Cosmology}
(Academic Press, Boston, 1990).
\item \label{b19} A.D. Linde, Phys. Lett. {\bf 175B} (1986) 395;
Physica
Scripta
{\bf T15} (1987) 169. 
\item \label{b20} 	A.S. Goncharov and A.D. Linde, Sov. Phys.
JETP {\bf
65} (1987) 635; A.S. Goncharov, A.D. Linde and V.F. Mukhanov, Int. J.
Mod.
Phys. {\bf A2} (1987) 561.
\item \label{b60} A.A. Starobinsky, in: {\bf Fundamental
Interactions}
(MGPI Press, Moscow, 1984), p. 55.
\item \label{b61} A.S. Goncharov and A.D. Linde, Sov. J. Part. Nucl.
  {\bf 17} (1986) 369.
\item \label{Star} A.A. Starobinsky, in: {\bf Current Topics in Field
Theory, Quantum Gravity and Strings}, Lecture Notes in Physics, eds.
H.J. de Vega and N. Sanchez (Springer, Heidelberg 1986) {\bf 206},
p. 107.
\item \label{HH} J.B. Hartle and S.W. Hawking, Phys. Rev. {\bf D28}
(1983) 2960.
\item \label{MezhMolch} A. Mezhlumian and S.A. Molchanov,  J. Stat.
Phys.,
{\bf 71} (1993) 799.
\item \label{ArVil} M. Aryal and A. Vilenkin, Phys. Lett. {\bf B199}
(1987) 351.
\item \label{Nambu} Y.Nambu and M.Sasaki, Phys.Lett. {\bf B219}
(1989) 240;
Y. Nambu, Prog. Theor. Phys. {\bf 81} (1989) 1037.
\item \label{Mijic} M. Miji\' c, Phys. Rev. {\bf D42} (1990) 2469;
Int.
J. Mod. Phys. {\bf A6} (1991)  2685.
\item \label{ZelLin} Ya.B. Zeldovich and A.D. Linde, unpublished
(1986).
\item \label{b99} T.S. Bunch and P.C.W. Davies, Proc. Roy. Soc. {\bf
A360}
(1978) 117; A. Vilenkin and L. Ford, Phys. Rev. {\bf D26} (1982)
1231;
A.D. Linde, Phys. Lett. {\bf 116B} (1982) 335;
A.A. Starobinsky, Phys. Lett. {\bf 117B} (1982) 175;
A. Vilenkin, Nucl. Phys. {\bf B226} (1983) 527.
\item \label{tunnel} A.D. Linde, JETP {\bf 60} (1984) 211; Lett.
Nuovo Cim.
{\bf 39} (1984) 401; Ya.B. Zeldovich and A.A. Starobinsky, Sov.
Astron.
Lett.{\bf 10} (1984) 135; V.A. Rubakov, Phys. Lett. {\bf 148B} (1984)
280; A.
Vilenkin, Phys. Rev. {\bf D30} (1984) 549.
 \end{enumerate}

\end{document}